\documentclass[aps,prl,twocolumn]{revtex4-1}

\usepackage{upgreek}
\usepackage{amsmath}
\usepackage{amsfonts}
\usepackage{amssymb}
\usepackage{graphicx}
\usepackage[colorlinks,linkcolor=blue,anchorcolor=blue,citecolor=blue,urlcolor=blue]{hyperref}
\usepackage{mathrsfs}
\usepackage{epsfig}
\usepackage{float}
\usepackage{subfigure}
\usepackage{mathtools}
\usepackage{url}
\usepackage{ulem}
\usepackage{verbatim}
\usepackage{color}

\begin{document}
\bibliographystyle{apsrev4-1}
\title{Conformal optical black hole for cavity}

\author{\mbox{Qingtao Ba$^{1,2}$, Yangyang Zhou$^{1,2}$, Jue Li$^{1,2}$, Wen Xiao$^{1,2}$, Longfang Ye$^{1,2}$, Yineng Liu$^{1,2}$}}
\author{Jin-hui Chen$^{1,2}$}
\email{jimchen@xmu.edu.cn}
\author{Huanyang Chen$^{1,2}$}
\email{kenyon@xmu.edu.cn}

\affiliation{$^{1}$Institute of Electromagnetics and Acoustics and Department of Physics, College of Physical Science and Technology, Xiamen University, Xiamen 361005, China,\\
$^{2}$Fujian Provincial Key Laboratory of Electromagnetic Wave Science and Detection Technology, Xiamen 361005, China.}

\date{\today}

\begin{abstract}
\noindent Whispering gallery mode (WGM) cavity is important for exploring physics of strong light-matter interaction. Yet it suffers from the notorious radiation loss universally due to the light tunneling effect through the curved boundary. In this work, we propose and demonstrate an optical black hole (OBH) cavity based on transformation optics.
The radiation loss of all WGMs in OBH cavity is completely inhibited by an infinite wide potential barrier. Besides, the WGM field outside the cavity is revealed to follow $1/r^\alpha$ decay rule based on conformal mapping, 
which is fundamentally different from the conventional Hankel-function distributions in a homogeneous cavity.
Experimentally, a truncated OBH cavity is achieved based on the effective medium theory, and both the Q-factor enhancement and tightly confined WGM field are measured in the microwave spectra which agree well with the theoretical results. The circular OBH cavity is further applied to the arbitrary-shaped cavities including single-core and multi-core structures with high-Q factor via the conformal mapping. The OBH cavity design strategy can be generalized to resonant modes of various wave systems, such as acoustic and elastic waves, and finds applications in energy harvesting and optoelectronics.
\end{abstract}
\maketitle

Whispering gallery mode (WGM) cavity is an intriguing platform for intensely enhancing light-matter interaction, which lays the foundations for ultra-low threshold lasers \cite{spillane2002ultralow,he2013whispering}, ultra-sensitive sensing \cite{zhu2010chip,foreman2015whispering,toropov2021review}, nonlinear optics \cite{zhang2019symmetry,lu2019efficient,chen2019microcavity} and quantum photonics \cite{haroche2006exploring,lodahl2017chiral,yang2021squeezed}. The conventional WGM cavity is composed of homogeneous materials with constant refractive index both in the core and cladding. The light field is confined in the cavity through the total internal reflection (TIR) and enhanced through constructive interference. The ultrahigh-Q factor has been realized in various dielectric WGM cavity with a large mode volume (V) and angular momentum \cite{lu2019efficient,zhang2019symmetry,zhang2017monolithic,lin2019broadband,wang2021high}. Nevertheless, the intrinsic radiation loss in an open boundary cavity with finite dielectric constant is ubiquitous, due to the light tunneling in the curved surface from the analog of quantum mechanics \cite{narimanov2009optical,xiao2010high,wyatt1962scattering}.
This radiation loss is remarkably increased and becomes the dominant loss mechanism when the resonant wavelength is comparable to the geometry size of the cavities \cite{spillane2005ultrahigh,matsko2006optical}. There is relentless effort for optimizing the Q/V in optical cavities, which is of great importance in exploring the cavity quantum electrodynamics (QED) \cite{vahala2003optical,buck2003optimal,spillane2005ultrahigh}. So far, various approaches have been proposed to manipulate the radiation loss and improve the Q-factor \cite{liu2016q,xiao2010high,jahani2014transparent,jiang2012highly}. For example, the plasmonic cavity \cite{xiao2010high,min2009high,chen2021recent} was constructed employing the strong optical-field localizations of metals, however, the strong ohmic loss in the plasmonic platform is unavoidable.  Alternatively, the radially anisotropic claddings were implemented to compress more energy into the core of cavity, resulting in tighter optical confinement and a substantially higher Q-factor \cite{liu2016q}. Unfortunately, for natural materials the anisotropic parameters are still challenging to implement. 
 
Transformation optics (TO) offers great versatility for manipulating light rays and electromagnetic
fields with novel functionalities in inhomogeneous dielectric materials  \cite{leonhardt2006optical,pendry2006controlling,chen2010transformation,lai2009illusion,xu2015conformal,kim2016designing} and structured metallic objects \cite{pendry2015transforming,li2016transformation,schurig2006metamaterial}. Many fascinating optical structures designed by TO, enables light deflection and trapping to mimick the cosmology effects \cite{sheng2013trapping,liu2010graded,cheng2010omnidirectional,chen2020transformation,xiao2021mimicking,mackay2010towards,sheng2018definite,genov2009mimicking}. In this work, we utilize TO theory to construct a class of optical black hole (OBH) cavities, including the single-core and multi-core cavity. The WGM field outside a circular OBH cavity is revealed to follow an unconventional $1/r^\alpha$ decay rule from conformal mapping. Employing the effective potential model, we strictly prove that the radiation loss of WGM in the OBH cavity can be completely inhibited thus the radiation Q-factor is infinite. Experimentally, we demonstrate both the Q-factor enhancement and tight field confinement of an OBH cavity compared with a homogeneous cavity in the microwave spectra. This  work  paved the way of surface field manipulation with conformal transformation, which can be generalized to resonant modes of various wave systems, such as acoustic and elastic waves, and finds applications in energy harvesting and optoelectronics.

\begin{figure}[t]
\centering
  \includegraphics[width=8.5cm]{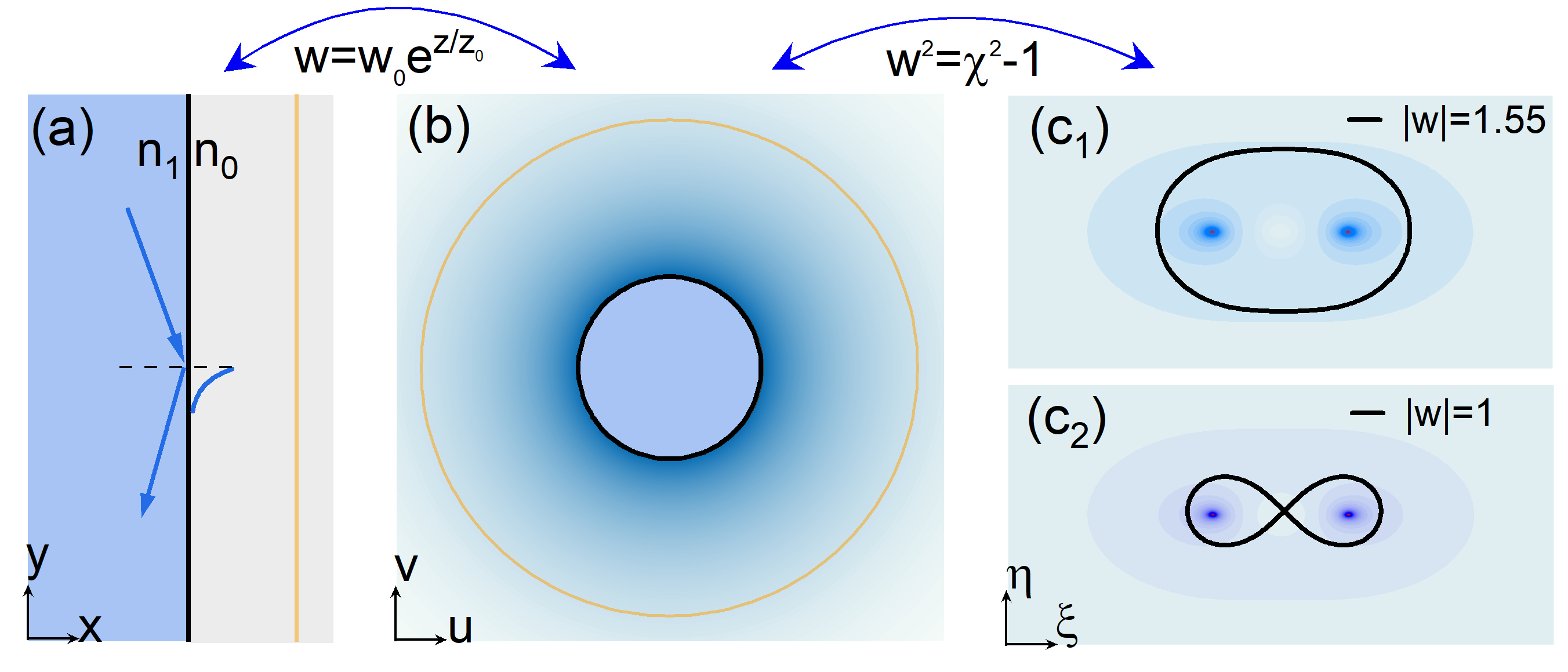}
   \caption{Conformal transformation for OBH cavity.
Mapping the homogeneous refractive index distribution in the original straight space (a) to a circular OBH cavity (b) with gradient index. The black and orange straight lines in (a) are transformed into the interior and exterior circle. The core region of OBH cavity is truncated as homogeneous index. The quadrupole cavity (c$_1$) and peanut-like cavity (c$_2$) is transformed from the circular OBH cavity under different constraint parameters. The black curve is the cavity boundary.}
  \label{fig:boat1}
\end{figure}

Let's start by reviewing the TIR at the planar interface between two dielectric media as shown in Fig.  \ref{fig:boat1}a, considering a light beam with incident angle $\theta_i$ impinging on the boundary from a denser medium ($n_1$). When $\theta_i$ is greater than the critical angle  $\theta_c$ ($\theta_c$=arcsin$(\frac{n_0}{n_1}))$, only the decaying evanescent field exists in the less dense medium ($n_0$). 
Based on the TO theory, wave propagation in original space with homogeneous index ($n_z$) could be equivalent to that in the physical space with inhomogeneous index distribution. As a result, we can construct a two-dimensional (2D) OBH cavity  in physical space ($w = u+iv$) by implementing an exponential transformation: $w = {w_0}{e^{z/{z_0}}}$ as shown in Fig. \ref{fig:boat1}b, where $z = x +iy$ is complex variables of the original space, $w_0$ and $z_0$  are the constants in $w$ space and $z$ space, respectively. 
The refractive index profile of OBH cavity is derived as:
\begin{equation}
n(r)=\left\{ \begin{array}{cl}
n_z\Big|\frac{\mathrm{d} z}{\mathrm{d} w}\Big|=\frac{n_0}{r/R} & r >R\\
n_1 &  r \leq R \end{array}\right.
\label{equ:1}
\end{equation}
Note that we only conformally map the index distribution in the claddings ($r>R$) and keep the core region ($r\leq R$) intact to avoid the divergence of index in the cavity center \cite{Supplemental}. 
As a consequence, the evanescent wave in $w$-space with polar coordinates can be obtained directly from the TIR-field transformation \cite{Supplemental}:
\begin{equation}
E(r,\varphi) =A{({\textstyle{R \over r}})^\alpha }{e^{im_w\varphi }}
\label{equ:ev}
\end{equation}%
where $A$ is field amplitude, $\alpha = \sqrt{m_w^2-{(n_0k_0R)^2}} $, $m_w = \beta R$, $\beta$ and $k_0$ are the propagation wave vector in $z$-space and vacuum respectively, $\varphi$ is the azimuthal angle. Therefore, the optical field in the cladding of the OBH cavity should follow the fast decay rule ($\sim 1/r^\alpha$), which is radically different from the Hankel-function distributions (asymptotic $\sqrt{1/r}$) of a homogeneous cavity.
Based on the form invariance of Maxwell’s equations under coordinate transformations, a group of conformal OBH cavity of single-core and multi-core shapes can be designed. For instance, we implement the mapping of ${w^2} = {\chi^2} - 1$ on circular OBH cavity [equation (\ref{equ:1})], where $\chi=\xi+i\eta$, and the quadrupole OBH cavity (peanut-like cavity) with $\left| w \right|$=1.55 ($\left| w \right|$=1) can be constructed as shown in Fig.  \ref{fig:boat1}c$_1$ (Fig. \ref{fig:boat1}c$_2$). The optical properties of these non-circular cavities are discussed later.

\begin{figure}[t]
\centering
  \includegraphics[width=8.5cm]{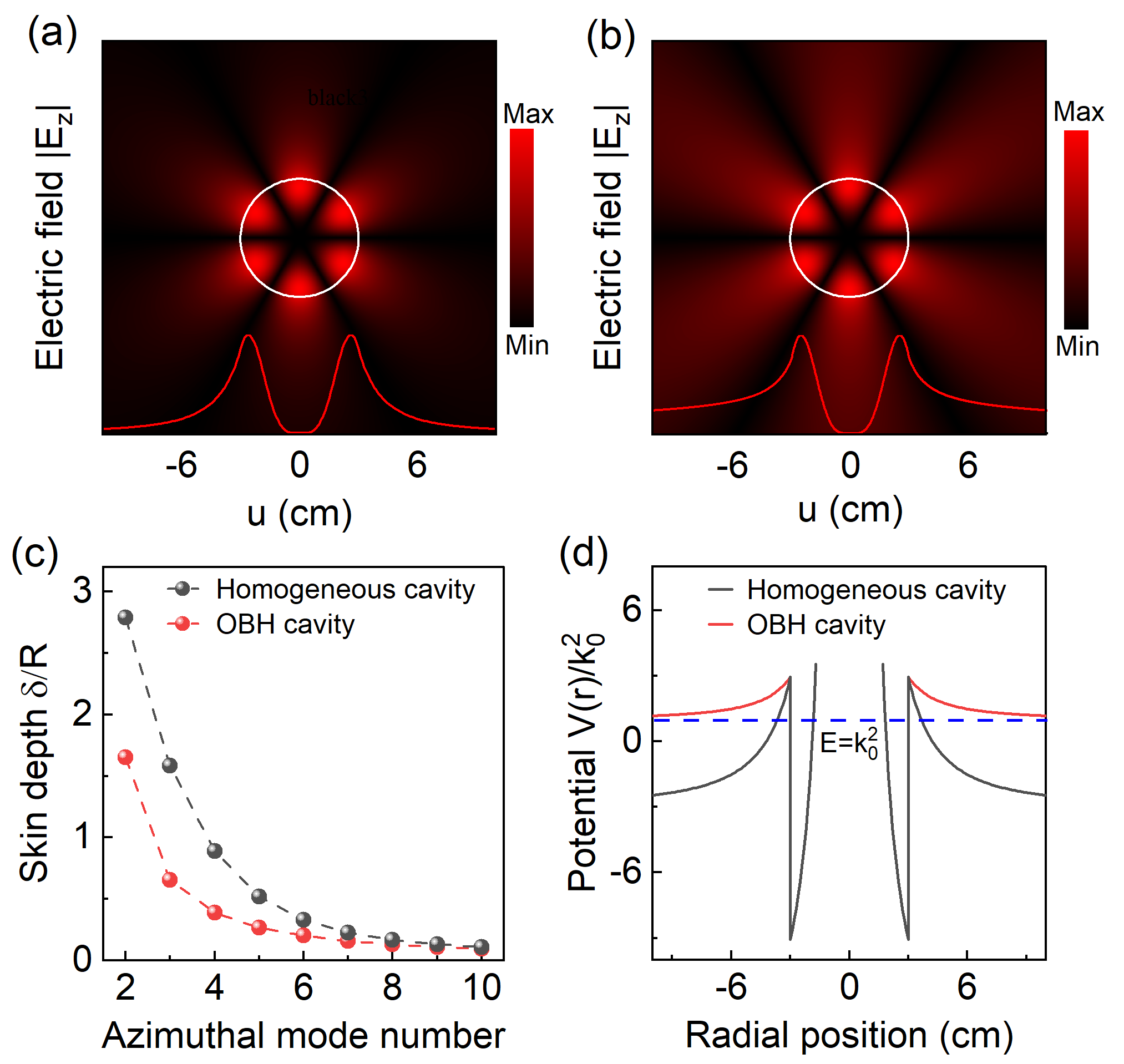}
  \caption{
Optical properties of WGM in OBH cavity and homogeneous cavity. Analytical standing wave pattern of WGM in OBH cavity (a) and homogeneous cavity (b), where the resonant mode-number $m$ is 3 and eigen-frequency is $\sim$ 1.93 GHz. Inset: normalized electric field amplitude in the radial direction. (c) Comparison of skin depth of WGM field in the OBH cavity and homogeneous cavity with various azimuthal mode numbers. (d) Effective potential of OBH cavity (black curve) and homogeneous cavity (red curve). }
  \label{fig:boat2}
\end{figure}

Now we analyze the WGM of circular OBH cavity via a direct solution of Maxwell's equations in the cylindrical coordinates. Since the transverse electric (TE) and transverse magnetic (TM) polarizations are decoupled, we consider the TE polarization without the loss of generality. Employing the method of separating variables with the boundary continuity, the WGM fields in the cladding region of OBH cavity is derived \cite{Supplemental}, which is exactly the same as equation (\ref{equ:ev}). The eigen-equation of OBH cavity is as following: 
\begin{equation}
{k_0}{n_1}R{J'_m}({n_1}{k_0}R) =-\sqrt {{m^2}-{{({n_0}{k_0}R)}^2}} {J_m}({n_1}{k_0}R)
\label{equ:2}
\end{equation}
where $m$ is an integer, $J_m$($n_{1}k_{0}R$) is the $m$-order Bessel function. It is found that 
equation (\ref{equ:2}) can have the real eigen-value under certain conditions \cite{Supplemental}, which is completely different from the homogeneous cavity with complex eigenfrequency. Thus the radiation loss caused by the curved boundary is completely inhibited in OBH cavity \cite{john2003optical}.

\begin{figure*}[htp]
  \centering
  \includegraphics[width=15.5cm]{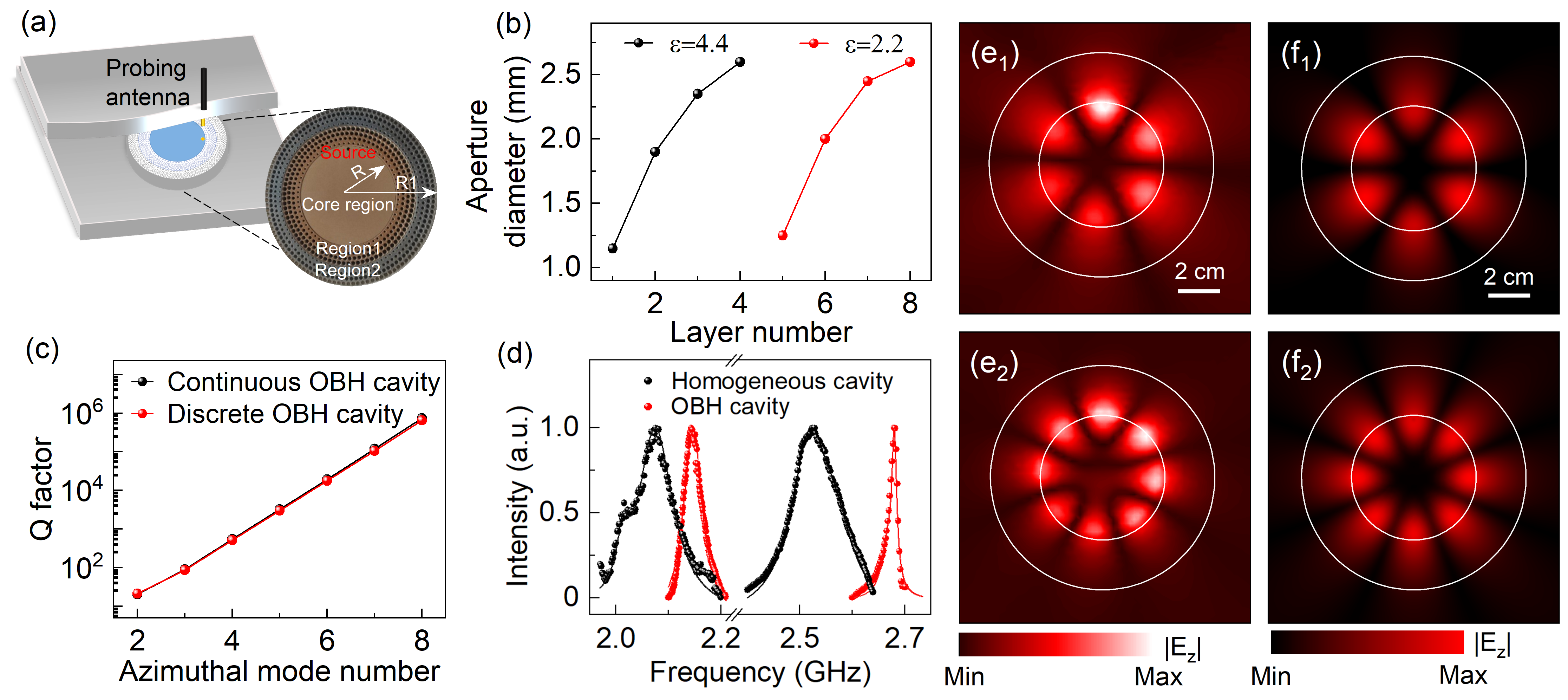}
\caption{Experimental characterizations of OBH cavity. (a) Experimental setup of the microwave near-field scanning system. Inset: photograph of the fabricated sample. (b) Design aperture size distributions in 8-layer structures. (c) Comparison of simulated Q-factor between continuous OBH and discrete OBH with various azimuthal mode numbers. (d) Experimental normalized resonant spectra of OBH cavity and homogeneous cavity with mode number $m$=3 and $m$=4. The solid curves are Lorentz fittings. Measured WGM field pattern with mode number $m$=3 (e$_1$) and $m$=4 (e$_2$). Simulated WGM field pattern with mode number $m$=3 (f$_1$) and $m$=4 (f$_2$). The white curves are the core and cladding material boundary of OBH cavity device.}
 \label{fig:boat3}
\end{figure*}

The analytical results of WGM fields in OBH cavity are shown in Fig. \ref{fig:boat2}a. Here, the structural parameters of circular OBH cavity are as following: $R$=3 cm, $n_1$=4, $n_0$=2, which facilitates experimental verification. In this study, we only focus on the lowest radial mode numbers, the optical fields of which are closely confined to the cavity boundary and have the smallest radial extent. As a comparison, the WGM fields of a homogeneous cavity with constant refractive index ($n_1'$=4, $n_0'$=2) is calculated as shown in Fig. \ref{fig:boat2}b. The WGM fields in the OBH cavity are tightly confined near the boundary, on the contrast, there is a considerable part of optical fields leaking into the cavity surroundings in the homogeneous cavity. To quantitatively characterize the field confinement, we define a skin depth (SD) $\delta$, i.e., the distance between the cavity boundary and the position with $1/e$ field-amplitude maximum. The SD of OBH cavity is always smaller than that of the homogeneous cavity as shown in Fig. \ref{fig:boat2}c. Since the optical field confinement is enhanced with the increase of resonant mode number, SD is decreased with the mode number as well. 

To more clearly elucidate the functions of the OBH refractive index profile, we modify the radial cylindrical Bessel equation as the analog of the Schrödinger equation \cite{johnson1993theory}. The central potential function $V (r)$ of light
propagation in OBH cavity is:
\begin{equation}
V(r) = k_0^2[1-n^2(r)]+\frac{(m+1/2)(m-1/2)}{r^2}
\label{equ:4}
\end{equation}
In particular, the potential function in the cladding region ($r>R$) is $V(r)$ = $k_0^2$+$({m^2}-{n_0^2}{k_0^2}{R^2}-1/4)/{r^2}$ as shown in Fig. \ref{fig:boat2}d. Obviously, the OBH cladding can relax the change of potential barrier, and the eigen-energy ($k_0^2$) of WGM in OBH cavity can be always smaller than the potential under the aforementioned cavity parameters \cite{Supplemental}. Thus the WGM field is tightly confined within the potential well, and the classical radiation loss is completely inhibited theoretically. On the contrast, in a homogeneous cavity, the light can always tunnel out the open boundary since the eigen-energy of WGM is larger than the effective potential when $r\gg R$ (Fig. \ref{fig:boat2}d black curve), and the radiation loss is not zero in all the WGMs. Note that in a conventional homogeneous cavity, when the cavity geometry is comparable to the light wavelength (resonant mode-number is small), the radiation loss is predominant and the Q-factor is greatly reduced. Although OBH cavity with a finite shell model has been theoretically studied to realize broadband light trapping/absorption and field penetration enhancement \cite{narimanov2009optical,zhu2012radially}, here we for the first time reveal that the total radiation inhibition with infinite radiation Q-factor can be achieved in an ideal OBH cavity structure.\\
\indent From equation (\ref{equ:1}), it shows that an ideal OBH cavity requires an infinite large region with gradient-index cladding ranging from $n_0$ to 0. This is challenging to realize in conventional materials although not impossible \cite{sheng2013trapping,jahani2016all,wang2017self,moitra2013realization}. Here, as an executable alternative, we reserve the range of cladding index from $n_0$ to 1, by truncating the region with index less than 1 in OBH cavity. Note that the index truncation breaks the ideal potential barrier, and it results in the tunneling leakage. Then, the continuous OBH cladding is discretized into 8 layers \cite{Supplemental}, and based on effective medium theory, the discrete layered index profile is achieved by drilling air holes with different diameters in two homogeneous dielectric slabs (Fig. \ref{fig:boat3}a-b). Here, three TF-2 dielectric plates with different permittivity are employed, which are 16, 4.4 and 2.2 from inside to outside. The effectiveness of this discontinuous multi-layer cavity structure is examined by numerical simulation as shown in Fig. \ref{fig:boat3}c. Finally, the fabricated sample, with 5.4 cm in radius and 2 cm in height (Fig. \ref{fig:boat3}a inset), were constructed by precision machining of dielectric plates. 

In microwave measurements, a point source was used to excite the WGM, and the resonant field was extracted by a near-field scanning system as shown in Fig. \ref{fig:boat3}a. The scanning system was controlled by a three-axis motion controller with a sampling step of 2 mm. As comparison, a homogeneous cavity including core region of index $n_1'$=4 and cladding region of index $n_0'$=2 (6 cm in radius and 2 cm in height) was fabricated and measured. From the resonant spectra of OBH cavity (blcak curve) and homogeneous cavity (red curve) in Fig. \ref{fig:boat3}d, the extracted Q-factor of WGM mode $m=3$ ($m=4$) is $\sim$ 49 ($\sim$ 241) for OBH cavity; as for homogeneous cavity, the Q-factor of WGM mode $m=3$ ($m=4$) is $\sim$ 25 ($\sim$ 30). The maximized Q-factor enhancement is $\sim$ 8 for mode $m$=4, which is smaller than the simulated Q-factor enhancement of $\sim$ 12. The discrepancy between experimental and theoretical results are caused by the strong light scattering in the drilling air hole and the limited precision of mechanical machining for OBH cavity sample. The measured field pattern around the resonant frequency for WGM modes ($m=3,4$) are plotted in Fig. \ref{fig:boat3}e, which agree well with the simulated results in Fig. \ref{fig:boat3}f. They unambiguously demonstrate the stronger field confinement than that of the homogeneous cavity \cite{Supplemental}. 

\begin{figure}[t]
  \includegraphics[width=8.5cm]{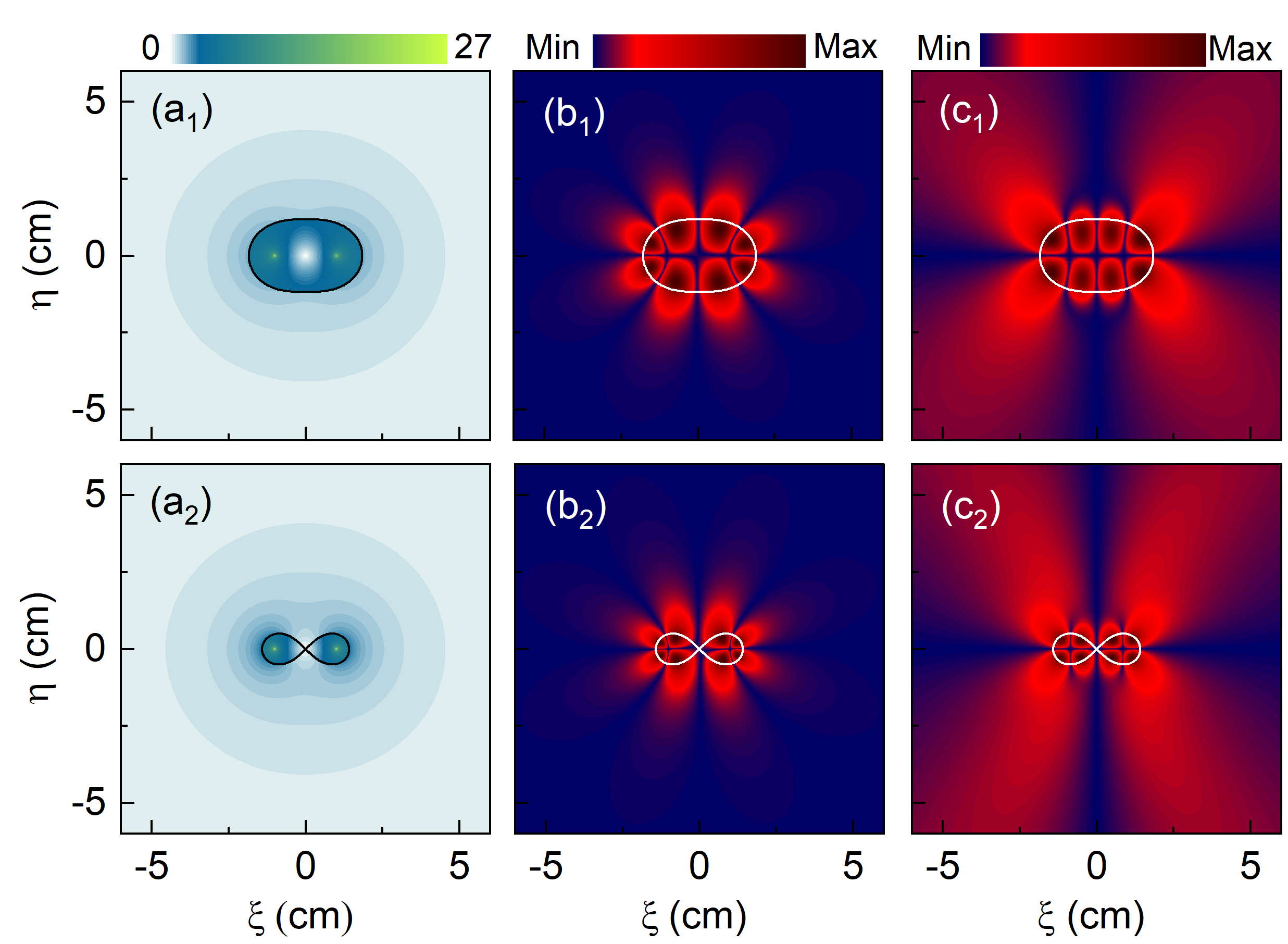}
  \centering
\caption{Comparison of WGM in the conformal OBH cavity and homogeneous cavity. The refractive index distribution for conformal quadrupole cavity (a$_1$) and peanut-like cavity (a$_2$), and their typical WGM field pattern with resonant mode-number $m$=4 (b$_1$-b$_2$). Typical WGM field pattern in a homogeneous quadrupole cavity (c$_1$) and peanut-like cavity (c$_2$). Black solid curve is the cavity boundary, the resonant frequency is $\sim$ 4.85 GHz and $\sim$ 7.3 GHz for quadrupole cavity and peanut-like cavity, respectively.
  }
  \label{fig:boat4}
\end{figure}

In addition to demonstrating radiation Q-factor enhancement and field confinement in a circular cavity, our designed device can be generalized to the deformed cavity for chaotic photon transport \cite{jiang2017chaos,chen2019regular}. For example, the face-shaped cavity and limaçon-shaped cavity with homogeneous core and OBH cladding shows stronger field confinement and Q-factor enhancement compared with a deformed homogeneous cavity \cite{Supplemental}. Furthermore, as previously suggested, using conformal mapping ${w^2} = {\chi^2}-1$ for circular OBH cavity, we have obtained quadrupole OBH cavity and peanut-like (dual cores) cavity with gradient index both in the core and cladding regions, as shown in Fig. \ref{fig:boat4}(a$_1$-a$_2$). It can be found that WGM fields in these conformal OBH cavity (Fig. \ref{fig:boat4}b) are tightly confined in the boundary compared with the homogeneous cavity (Fig. \ref{fig:boat4}c). Besides, the Q-factor of conformal quadrupole cavity (peanut-like cavity) is boosted by 6 (4) orders of magnitude compared with a homogeneous one. Given the ingenious conformal transformation, theoretically the OBH cavity with arbitrary boundary shapes can be constructed, which may provide a new avenue to cavity optics for fundamental physics and photonic applications.

In summary, we have reported a group of conformal OBH cavity based on transformation optics. The universal radiation loss in a conventional homogeneous cavity is completely inhibited in the OBH cavity by an infinite wide potential barrier. The WGM field in OBH cavity is revealed to follow an unconventional $1/r^{\alpha}$ decay rule based on the conformal mapping. We demonstrate both the Q-factor enhancement and strongly confined WGM field for a truncated OBH cavity in the microwave spectra. Furthermore, we show that the designed OBH cavity device of circular shape can be applied to the arbitrary-shaped cavity, which might find applications for chaotic photon transport with high-Q factor. Although we focus on the WGM in electromagnetic waves, it is promising to extend the spectra into the optical frequency band for cavity QED and optoelectronics. 

\begin{acknowledgments}
This research was funded by National Natural Science Foundation of China (62005231, 92050102); National Key Research and Development Program of China (2020YFA0710100). Q. B. and Y. Z. contributed equally to this work. J.-h. C. and H. C. conceived and supervised the project. All authors contributed to the discussion, analyzed the data, and wrote the manuscript. We would also like to thank Pengfei Zhao and Qi-Tao Cao for helpful discussion.
\end{acknowledgments}

\normalem
\bibliography{REF.bib}

\end{document}